\documentstyle[preprint]{ptptex}
\newcommand{\bl}[1]{\mbox{\boldmath{$ #1$}}}
\newcommand{\blsmall}[1]{\mbox{\boldmath{ \footnotesize $ #1$}}}
\newcommand{\epsl}{\varepsilon}
\newcommand{\f}[2]{\frac{#1}{#2}}
\newcommand{\dd}{\partial}

\preprintnumber[3cm]{KNK-01I22}

\title{%        %You can use \\ for explicit line-break
Statistics of the Composite Systems
}
\author{%       %Use \sc for the family name
Hitoshi {\sc Ito}\footnote{
 E-mail address: itoh@phys.kindai.ac.jp}
}

\inst{%         %Affiliation, neglected when [addenda] or [errata]
Department of Physics, Faculty of Science and
 Engineering,\\ Kinki University, Higashi-Osaka, 577 }

\recdate{%      %Editorial Office will fill in this.
}
\date{June 26, 2001}

\abst{%       %this abstract is neglected when [addenda] or [errata]
The commutation relations of the composite fields are studied in the 3, 2 and 1 space dimensions. It is shown that the field of an atom consisting of a nucleus and an electron fields satisfies, in the space-like asymptotic limit, the canonical commutation relations within the sub-Fock-space of the atom. 
The composite anyon fields are shown to satisfy the proper anyonic commutation relations with the additive phase exponents. Then, (quasi)particle pictures of the anyons are clarified. The hierarchy of the fractional quantum Hall effect is rather simply understood by utilizing the (quasi)particle charactors of the anyons. 
The commutation relations of the scalar object in the Schwinger(Thirring) model are mentioned briefly.
}

\begin{document}

\maketitle

\section{Introduction}

More than forty years ago, composite systems in the quantum field theory was investigated\cite{Zimmermann58}. 
One starts from a local scalar field, say $A(x)$, and assumes the existence of discrete eigenvalues $m^2$ and $M^2$ of the 4-momentum squared $P^2$, where $m$ is the mass of the original field $A$.
If $\langle A(x)A(y)\mid P^2=M^2\rangle \neq 0$ there may be a composite(bound)  state of the mass $M$. Then, one defines a bilocal field
\begin{equation}
  B(x,\epsl)=TA(x+\epsl)A(x-\epsl)        \label{a1}
\end{equation}
representing this state, where $T$ denotes the time-ordered product.
Zimmermann showed, under some mathematical assumption, that the asymptotic($t\to\pm\infty$) field of $B$ satisfies the proper commutation relation in the limit $\epsl\to 0$, if it is suitablly normalized.
 One can infer, from this construction, that composite systems can be described  by the local field operators and there are no differences between elementary and composite particles in constructing the $S$ matrix elements.
An idea of the boot strapping(nuclear democracy) emerged from this observation.

However, the successes of the gauge theories have changed drastically the framework of the elementary particle theory. 
The quantum theory of field has revived and one gradually recognized the hierarchy structure of the gauge interactions. 
A field theory now becomes an effective theory for each class of the hierarchy.

On the other hand, there is another hierarchy of compositeness in nature\cite{Sakata61}, which was revealed through success of the composite models of elementary particles. The hierarchy here consists of the classes of quark, hadron, neucleus, atom and so on.
 The level of a class is specified by the energy scale indicating the limit of applicability of the theory governing it.
We, further, believe that the theory is a quantum field theory and there exist elementary fields for each class, which are constructed from the elementary fields of the deeper class.
The most instructive example of this interpolating mechanism may be provided by considering the class of atom. 
An atom is a composite system which consists of a nucleus field and an electron field interacting through the photon field. Then, we construct the atom field as a composite field of the elementary fields of the deeper class. In this respect, we are specially interested in the statistic property of the composite system, since we are hardly convinced of it by the particle quantum mechanics:
 For example, the nucleus of the hydrogen atom is a Fermi particle and it changes into the Bose particle by acquiring another fermion. This is mysterious from the view point of the particle quantum mechanics\footnote{
The gas of the H-atoms undergoes the Bose-Einstein condensation. It means that the protons distribute like Bose particles, since the position of them almost coincide with those of the atoms. 
}.

The difference between the quantum field theory and the many-particle quantum mechanics is more apparent in the world of the 2 space dimension. Its topological structure allows the exotic statistics of the field operators, which cannot always be realized by the flux-charge composite having the characteristics of (quasi)particle.

We study the field aspects of the hydrogen atom in \S 2 and show the canonical commutation relations for the atom field. \S 3 is devoted to studying the anyon fields in the space dimension of 2. We show the additivity of the phase exponents in the commutation relations and study some case of the anyons having the charactors of (quasi)particles. Some aspects of the phenomenology in the fractional quantum Hall effect(FQHE) are sketched on this basis. In the 1 space dimension, composite field has rather clear-cut structure of the commutation relations, which is discussed in \S 4. We discuss conceptual sides of the subject in the last section.

\section{The atom field  \label{atom}}

Let us first introduce the composite field $\Psi$ through the equation
\begin{equation}
   \Psi(x_1,x_2)=T\psi(x_1)\phi(x_2)       \label{a2}
\end{equation}
where $\psi(x_1)$  and $\phi(x_2)$ are the elementary fields of the nucleus and the electron respectively in the Heisenberg picture. We, then, define the Bethe-Salpeter amplitude $\langle 0|\Psi(x_1,x_2)|2\rangle$ for the two-particle states and obtain the state of the H-atom by solving the BS equation for it. The nonrelativistic approximation suffices for the present purpose. And we are interested in only the wave function of the ground state, the Fourier transform of which is denoted by $g_{mm'}(\bl{k}_1,\bl{k}_2)$, where $m$ and $m'$ are the indices of the spin of the nucleus and the electron respectively. We note, however, that these indices are dummy since the spin of the nucleus is frozen in the nonrelativistic situation and therefore the spin freedom of the electron can be neglected in the ground $S$ state. 

We next reconstruct the composite field operator by including the bound-state amplitude. If the total and the relative momenta are $K=(K_0,\bl{K})$ and $k=(k_0,\bl{k})$ respectively, the contribution of the bound state to the annihilation part is given by
\begin{eqnarray}
 \lefteqn{ \Psi(X,x) = Ce^{iKX}\sum_{mm'}\int d^3k} \label{a3} \\
   & & \times g_{mm'}(\bl{K},\bl{k})
    a_m(\eta_1\bl{K}-\bl{k})b_{m'}(\eta_2\bl{K}+\bl{k})
                            \exp{(i\bl{k}\cdot\bl{x})},
                     \quad \eta_1+\eta_2=1,      \nonumber
\end{eqnarray}
where $C$ is a normalization factor and $a_m(\bl{k}_1)$ and $b_{m'}(\bl{k}_2)$ are the annihilation operators respectively for the nucleus and the electron which satisfy the anti-commutation relations
\begin{equation}
    \{a_m(\bl{k}_1),a^\dagger_{m'}(\bl{k}_2)\}=
      \delta_{mm'}\delta(\bl{k}_1-\bl{k}_2), \quad \mbox{etc.}.   \label{a4}
\end{equation}
Now, when we observe the stable atom from some great distance we can neglect the scale of the relative coordinate $\bl{x}$ and the atom is represented by the wave function at the origin
\footnote{
We should use a relativistic equation in deeper classes of the compositeness hierarchy. Then, the wave function at the origin becomes a divergent quantity in some cases. We have to renormalize it\cite{Ito82}.
}.
 We call, after Haag, the neglect of the relative coordinate the space-like asymptotic limit\cite{Zimmermann58}.

The bound state is represented by a local field operator in the space-like asymptotic limit, the annihilation part of which is given by
\begin{equation}
  \Psi(X,0)=CA(\bl{K})e^{iKX},  \quad 
   A(\bl{K})=\int g(\bl{K},\bl{k})a(\eta_1\bl{K}-\bl{k})b(\eta_2\bl{K}+\bl{k})d^3k,  
                               \label{a5}
\end{equation}
where the dummy spin indices are omitted. $A(\bl{K})$ satisfies commutation relations
\begin{equation}
   [A(\bl{K}),A(\bl{K}')]=[A^\dagger(\bl{K}),A^\dagger(\bl{K}')]=0  \label{cc1}
\end{equation}
and
\begin{eqnarray}
 \lefteqn{ [A(\bl{K}),A^\dagger(\bl{K}')] = \delta(\bl{K}-\bl{K}')} 
      \label{cc2} \\ & & 
   -\int d^3k g(\bl{K},\bl{k})g^*(\bl{K}',\eta_2\bl{K}-\eta_2\bl{K}'+\bl{k})
 a^\dagger(\bl{K}'-\eta_2\bl{K}-\bl{k})a(\eta_1\bl{K}-\bl{k})  \\ & &
 -\int d^3k g(\bl{K},\bl{k})g^*(\bl{K}',\eta_1\bl{K}'-\eta_1\bl{K}+\bl{k})
 b^\dagger(\bl{K}'-\eta_1\bl{K}+\bl{k})b(\eta_2\bl{K}+\bl{k}) \nonumber
\end{eqnarray}
where the normalization of the mometum-space wave function is assumed to be 1. The spectral condition forbids the last two terms in the right hand side to have matrix elements within the subspace of the bound state. We therefore neglect them and reach the canonical commutation relations for the asymptotic atom field. 

\section{Anyon fields in the 2 space dimension}

The anyon is a particle-like excitation in the 2+1 dimensional system, which is observed, for example, in the phenomena of the FQHE. It is characterized by the fractional statistics in which interchange of two anyons results in any phase of the wave function\cite{Wilczek82}. 
The electrons system in the FQHE is confined in some 2 dimensional surface by the complicated electromagnetic interactions with the surrounding materials. The most important effect of the confinement is a change of the topological structure of the configuration space. In the 2+1 dimensional field theory, this boundary condition is settled by allowing the Chern-Simons(CS) gauge term in the Lagrangian, which is a mathematical device to replace the confining interaction and leads to the exotic statistics\cite{Semenoff88}.

\subsection{Exotic statistics in the Chern-Simons field theory}

We consider two species of the charged particles the fields of which are denoted by $\psi$ and $\phi$. We assume, for definiteness, the bosonic cmmutation relations among them and further introduce three CS terms in the Lagrangian according to Ezawa-Hotta-Iwazaki\cite{Ezawa92}\cite{Wilczek92}. Then, the CS part of the Lagrangian becomes

\begin{eqnarray}
\mathcal{L}_{\small \mbox{CS}} & = &
    (\dd_\mu+ia_\mu)\psi^*(\dd^\mu-ia^\mu)\psi-m^2\psi^*\psi
  -\f{1}{4\alpha}\varepsilon^{\mu\nu\lambda}a_\mu\dd_\nu a_\lambda \nonumber \\
       &  & (\dd_\mu+ib_\mu)\phi^*(\dd^\mu-ib^\mu)\phi-M^2\phi^*\phi
  -\f{1}{4\beta}\varepsilon^{\mu\nu\lambda}b_\mu\dd_\nu b_\lambda \nonumber \\
        & & -\f{1}{4\gamma}\varepsilon^{\mu\nu\lambda}(a_\mu\dd_\nu b_\lambda +
                           b_\mu\dd_\nu a_\lambda),     \label{Lcs}
\end{eqnarray}
where the last term governs the mutual statistics between the fields $\psi$ and $\phi$. We quantize this system by following the procedure developed by Semenoff\cite{Semenoff88}\cite{Matsuyama89}. The constraint conditions become

\begin{eqnarray}
 \f{1}{2\alpha}\varepsilon_{ij}\dd_ia_j+\f{1}{2\gamma}\varepsilon_{ij}\dd_ib_j,
        & = & j_0,                \label{consta} \\
 \f{1}{2\gamma}\varepsilon_{ij}\dd_ia_j+\f{1}{2\beta}\varepsilon_{ij}\dd_ib_j 
        & = & k_0 ,
\end{eqnarray}
where $j_0$ and $k_0$ are the 0th component of the current of $\psi$ and $\phi$ respectively. Under these conditions we get the Hamiltonian in which $\psi$ and $\phi$ couple minimally to $a_i$ and $b_i$(i=1,2). Further, by assuming the gauge conditions $\dd_ia_i=\dd_ib_i=0$ we find that the CS fields are given by

\begin{eqnarray}
 a_i(x) & = & \f{1}{\pi}\dd_i\int d^2y\Omega(\bl{x}-\bl{y})
           \{\mu_aj_0(y)-\mu k_0(y)\}\equiv \dd_i\Theta_a(x), \label{ai} \\
 b_i(x) & = & \f{1}{\pi}\dd_i\int d^2y\Omega(\bl{x}-\bl{y})
           \{\mu_bk_0(y)-\mu j_0(y)\}\equiv \dd_i\Theta_b(x), \label{bi} \\
 \Omega(\bl{x}-\bl{y}) & = & \mbox{arctan} \f{x^2-y^2}{x^1-y^1}.
\end{eqnarray}
The coefficients $\mu$'s are given by

\begin{equation}
 \mu_a=\f{\alpha\gamma^2}{\gamma^2-\alpha\beta},\quad
\mu_b=\f{\beta\gamma^2}{\gamma^2-\alpha\beta},\quad
  \mu=\f{\alpha\beta\gamma}{\gamma^2-\alpha\beta}.
\end{equation}

The equations (\ref{ai}) and (\ref{bi}) suggest that we can eliminate the CS gauge interactions from the Hamiltonian by applying the (singular)gauge transformations

\begin{equation}
 \psi(x)=\exp(i\Theta_a(x))\hat{\psi}(x), \quad 
    \phi(x)=\exp(i\Theta_b(x))\hat{\phi}(x) \quad \mbox{etc.}. \label{anytra}
\end{equation}
Then, the equal-time commutation relations become exotic after this singular transformations:

\begin{eqnarray}
\hat{\psi}(x)\hat{\psi}(y)-e^{i(2n+1)\mu_a}\hat{\psi}(y)\hat{\psi}(x) &=&0,
                                                            \nonumber \\
\hat{\psi}(x)\hat{\pi}(y)-e^{i(2n+1)\mu_a}\hat{\pi}(y)\hat{\psi}(x)
                   &=& i\delta(\bl{x}-\bl{y}),  \nonumber \\
\hat{\phi}(x)\hat{\phi}(y)-e^{i(2n'+1)\mu_b}\hat{\phi}(y)\hat{\phi}(x) &=&0,
                                                            \nonumber \\
\hat{\phi}(x)\hat{\chi}(y)-e^{i(2n'+1)\mu_b}\hat{\chi}(y)\hat{\phi}(x)
                   &=& i\delta(\bl{x}-\bl{y}),     \nonumber \\
\hat{\psi}(x)\hat{\phi}(y)-e^{i(2n''+1)\mu}\hat{\phi}(y)\hat{\psi}(x) &=&0,
                                                       \nonumber \\
     & \mbox{etc.}, &               \label{psiphi}
\end{eqnarray}
where $\pi$ and $\chi$ are the fields canonical conjugate to $\psi$ and $\phi$ respectively. The odd integer $2n+1$, etc. come from the multi-valuedness of the function $\Omega$.

\subsection{Composite anyon field \label{company}}

If we express the free anyon fields as superpositions of the plane waves

\begin{eqnarray}
 \hat{\psi}(x) & = & \int\f{d^2k}{2\pi}\f{1}{2\omega}
           \{a(\bl{k})e^{-ikx}+c^\dagger(\bl{k})e^{ikx}\},  \\
 \hat{\phi}(x) & = & \int\f{d^2k}{2\pi}\f{1}{2\omega}
                 \{b(\bl{k})e^{-ikx}+d^\dagger(\bl{k})e^{ikx}\},
\end{eqnarray}
the commutation relations among the operators $a$, $b$, $c$... becomes

\begin{eqnarray}
 a(\bl{k})a(\bl{k}')-e^{i(2n+1)\mu_a}a(\bl{k}')a(\bl{k}) &=& 0 \nonumber \\
 a(\bl{k})a^\dagger(\bl{k}')-e^{i(2n+1)\mu_a}a^\dagger(\bl{k}')a(\bl{k})
                            & = & \delta(\bl{k}-\bl{k}') \nonumber \\
 b(\bl{k})b(\bl{k}')-e^{i(2n'+1)\mu_b}b(\bl{k}')b(\bl{k}) & = & 0
                                                        \nonumber \\
  b(\bl{k})b^\dagger(\bl{k}')-e^{i(2n'+1)\mu_b}b^\dagger(\bl{k}')b(\bl{k})
                            & = & \delta(\bl{k}-\bl{k}') \nonumber \\
 a(\bl{k})b(\bl{k}')-e^{i(2n''+1)\mu}b(\bl{k}')a(\bl{k}) &=& 0 \nonumber \\
                   & \mbox{etc.} & .  \label{abcom}
\end{eqnarray}

When $\mu'\mbox{s}/\pi$ are integers we can construct the Fock space and have the particle picture of anyons. We note that the condition by which anyons have the particle interpretation can be reluxed some what as it is discussed in \S \ref{anypart} and \S \ref{FQHE}.

%\subsection{Composite anyon field }

Let us assume that two anyon fields with the statistical parameters $\alpha$ and $\beta$ form a composite system. Relying on 'the glue is unimportant' principle, we can calculate the commutators of the composite operators in the same way as in \S \ref{atom}, where the anticommutators are replaced by the exotic ones (\ref{abcom}):
 The composite operator $A$ in the space-like asymptotic limit is defined by
 
 \begin{equation}
      A(K)=\int g(K,k)a(\eta_1\bl{K}-\bl{k})b(\eta_2\bl{K}-\bl{k}) d^2k.
 \end{equation}
$A$ satisfies, under the same consideration discussed in \S \ref{atom}, the commutation relations

\begin{eqnarray}
  A(\bl{K})A(\bl{K}')-e^{i(2n+1)\mu_{ab}}A(\bl{K}')A(\bl{K}) & = & 0,  \\
   A(\bl{K})A^\dagger(\bl{K}')-e^{i(2n+1)\mu_{ab}}A^\dagger(\bl{K}')A(\bl{K})
    & = & \delta(\bl{K}-\bl{K}'),
\end{eqnarray}
where\cite{endnote}

\begin{equation}
  \mu_{ab}=\mu_a+\mu_b, \quad \mbox{mod}\, 2\pi.      \label{muab}
\end{equation}

\subsection{Particle picture of the anyon    \label{anypart}}
    
We have assumed that the basic fields $\psi$ and $\phi$ have one unit of the CS(statistical) charge in the Lagrangian (\ref{Lcs}). The CS flux is, therefore, quantized in the unit $2\pi$
\footnote{We are using the natural unit $\hbar=1$.}. 
Suppose an object(anyon) a with the CS charge $m_a$ carrying the $f_a$ units of the flux which is perpendicular to the confined surface. If two of such object interchange their position by the half circular motion around each other, the two-particle wave function aquires the phase $\exp(\pm if_am_a\pi)$ by virtue of the Aharonov-Bohm effect for the CS flux. By comparing with (\ref{psiphi}), we are tempted to identify $f_am_a$ with $\mu_a/\pi$ . However, it is shown to be not always possible by inspecting the composite anyon state: It is natural to assume the conservations of charge and flux in the process of the binding. Then, the charge and the flux of which $\mu_{ab}$ consists are the sums of those possessed by the anyons a and b. But this assignment is not always compatible with (\ref{muab}) as is shown in the following.

We first note that anyonic particles cannot even coexist peacefully if they have different charge/flux ratios, since we cannot define the interchange itself of two anyons consistently for them. Consider, then, $n$ anyons with the flux $f_i$ and the charge $m_i$ which satisfy the relations

\begin{equation}
 m_i=k f_i, \quad i=1,2,...,n
\end{equation}
with a common constant $k$. Suppose next that these $n$ anyons form a composite state, which have the flux $\sum f_i$ and the charge $\sum m_i$. The equation (\ref{muab}) then gives the consistency condition

\begin{equation}
 k\{(\sum_{i=1}^nf_i)^2-\sum_{i=1}^nf_i{}^2\}=2p,
                                      \quad p=\mbox{integer}.  \label{sumcon}
\end{equation}
For the identical $n$ anyons this condition becomes $n(n-1)f_1m_1=2p$.

 The anyon does not have the particle picture in general. But, it can be regarded as a particle having the flux $f_L$ and the charge $m_L$ if the converted statistical parameter $\mu_L$ satisfies
 
\begin{equation}
  \f{\mu_L}{\pi}=p(\mbox{integer})=f_Lm_L,
\end{equation}
because we can construct the Fock space for this value of the phase. Now, suppose that this anyonic particle decays into $n$ anyons satisfying the cliterion (\ref{sumcon}). We can also regard these products of dissociation to be the psudo-particles, since their parent state has the particle picture.

\subsection{Anyons in the fractional quantum Hall effects  \label{FQHE}}

The phenomenology of the FQHE becomes simple by using the (quasi)particle concepts of anyons. We sketch it in this and the next subsections. 
The ground state(Laughlin state) of the FQHE is an incompressible quantum fluid made of the composite bosons with the charge $-e$ carrying $f_L$ units of the real magnetic flux\cite{Laughlin83}, where the quantization unit is $\phi_0=2\pi/e$. Since it is an anyonic particle its statistical parameter $\alpha_L$ is given by

\begin{equation}
 \f{\alpha_L}{\pi}=f_L=\f{B}{\phi_0\langle\rho\rangle_L},
                                                       \label{alphl}
\end{equation}
where $\langle\rho\rangle_L$ is the boson density and $B$ is the external magnetic field. We see that the filling factor $\nu^L$ is $1/f_L$ in this state. The CS field associated with the boson is given through (\ref{consta}) with $\alpha=\alpha_L$, $\gamma=\infty$ and $j_0=\rho$, from which we have 

\begin{equation}
\langle\rho\rangle_L=\f{1}{2\alpha_L}\varepsilon_{ij}\dd_i\langle a_j\rangle_L,
   \quad \alpha_L=\f{\pi B}{\phi_0\langle\rho\rangle_L}.
\end{equation}
We therefore have $\langle a_i\rangle_L=eA_i$ for the ground Laughlin state, where $\varepsilon_{ij}\dd_iA_j=B$.

Suppose that an elementary topological vortex is excited in the increasing magnetic field $B$\cite{Ezawa92}. It is given by the configuration

\begin{equation}
 \rho=\langle\rho\rangle_L+ \rho_q, \quad a_i=eA_i+v_i, \quad
                                 v_i \to\dd_i\theta(\mbox{at the infinity}),
\end{equation}
where $\theta$ is the azimuthal angle in the frame whose origin is at the center of the vortex. The CS flux of this excitation becomes $2\pi$, which amounts to a unit flux $\phi_0$ of the real magnetic field. On the other hand, the statistical charge $Q$ of the field $q$ is given by

\begin{equation}
 Q=\int\rho_qd^2x=\f{1}{2\alpha_L}\int\varepsilon_{ij}\dd_iv_jd^2x
                     =\f{\pi}{\alpha_L}=\f{1}{f_L}.
\end{equation}
Thus, the vortex is identified with the quasihole(anti-quasiparticle) having the unit flux and the CS charge $m_q=1/f_L$. We, further, find that the statistical parameter of the field $q$ is given by $\alpha_q=m_q\pi$ because the anyon $q$ is a dissociation product of the composite boson.

The accumulated quasiholes form the second incompressible-fluid state, where the next vortices are created, which again form the higher Laughlin state, and so on. Though the anyons in this hierarchical structure cannot be regarded as the products of the simple dissociation, they are the quasiparticle with the definite flux and the CS charge as is shown in the next subsection. 

\subsection{Hierarchy of the FQHE}

Let us begin with the $s$th quasiparticle of the density $\rho_s$ which has the charge $q_se$ and the statistical parameter $\alpha_s$. The quasiparticles form the Laughlin state with the density $\langle \rho\rangle_{s+1}$. Then, the $s+1$th vortex is excited, which is given by the configulation

\begin{eqnarray}
 \rho_s=\langle \rho\rangle_{s+1}+ \rho_{s+1}, \quad & a_{i(s)} & 
                      =\langle a_i\rangle_{s+1} +v_{i(s+1)},   \label{vortex}\\
        & v_{i(s+1)} & \to \tau_{s+1}\dd_i\theta\mbox{(at the infinity)},
\end{eqnarray}
where the vortex with $\tau_{s+1}=1(-1)$ is identified with the quasihole(quasiparticle). We assume that the frozen part $\langle\rho\rangle_{s+1}$ satisfies

\begin{equation}
 \langle\rho\rangle_{s+1}=
      \f{1}{2\alpha_s}\varepsilon_{ij}\dd_i\langle a_j\rangle_{s+1}.
\end{equation}
Then, the constraint equation gives

\begin{equation}
 \rho_{s+1}=\f{1}{2\alpha_s}\varepsilon_{ij}\dd_iv_{j(s+1)}.
\end{equation}
Now, the $s+1$th CS charge is given by

\begin{equation}
 m_{s+1}=\int \rho_{s+1}d^2x=\tau_{s+1}\f{\pi}{\alpha_s}.
\end{equation}
The parameter $m_{s}$ determines the statistics, which is also governed by the parameter $\alpha_s$ through the relations (\ref{psiphi}) with $\gamma=\infty$. We, therefore, identify $\alpha_s/\pi$ with $m_{s}$ and get the reciprocal relation\cite{Ezawa92}

\begin{equation}
 m_{s+1}=\f{\tau_{s+1}}{m_s}+2p_{s+1},   \label{recm}
\end{equation}
where $p_{s+1}$ is an integer representing the multivalued nature of the statistical factor.

There act the Coulomb forces among the quasiparticles of (\ref{vortex}) and the lowest energy state becomes the incompressible fluid. In order to estimate the charge $q_{s+1}$ of the anyons, we next investigate the structure of this fluid state\cite{Ando93}: The $N$ quasiparticles build up the $s+1$th Laughlin fluid, the wave function of which is given by

\begin{equation}
 \Psi_{s+1} \propto \prod_{i<j}(z_i-z_j)^{m_{s+1}}
            \exp(-\sum_{i=1}^N\f{|z_i|^2}{4\ell^2}),   \label{Laugh}
\end{equation}
where $z_i=x_i+iy_i$ and the index $m_{s+1}$ comes from the anyonic statistics. $1/2\pi\ell^2$ is the state density of the quasiparticle, which is $|q_s|$ times that of the electron. In the wave function (\ref{Laugh}) $M=N m_{s+1}$ is the highest angular momentum of the constituent quasiparticle and the area occupied by the system is given by $2\pi\ell^2Nm_{s+1}$. Now, the buried quasiparticle shows up itself as the excitation in this fluid. The quasihole excitation is expressed by the operation $S(z_0)$ on the wave function, since it has an unit of the CS flux;

\begin{equation}
 S(z_0)\Psi_{s+1}\propto \prod_i(z_i-z_0)\Psi_{s+1}.
\end{equation}
The area of the system increase by $2\pi\ell^2$ by this operation, which corresponds to the $1/m_{s+1}$ of the constituent $s$th quasiparticle. Since the quasihole excitation has the charge opposite to the constituent, we have the charge relation

\begin{equation}
 q_{s+1}=-\tau_{s+1}\f{q_s}{m_{s+1}}.    \label{recq}
\end{equation}

Getting back to (\ref{Laugh}), we next note that the number of the one-quasihole states is given by $Nm_{s+1}$ for the large $N$. Since the constituent quasiparticle has the charge $q_se$, the number of the electron-equivalent states(the number of the flux quanta) becomes $Nm_{s+1}/|q_s|$. On the other hand, since the total charge of the quasi-particles is $\tau_{s+1}Nq_se$, the charge fraction per the electron-equivalent state is $\tau_{s+1}q_s|q_s|/m_{s+1}$. By considering the change in the charge fraction we have the recursion formula for the filling factor $\nu^s$

\begin{equation}
 \nu^{s+1}=\nu^s -\tau_{s+1}\f{q_s|q_s|}{m_{s+1}}.  \label{recf}
\end{equation}

We finally obtain the hierarchy of the FQH states from (\ref{recm}), (\ref{recq}) and (\ref{recf}).\cite{Halperin84}

We have assumed the anyon transformation (\ref{anytra}) and eliminated the CS fields at every lebels of the hierarchy. Another way to get the hierarchy of the filling factors is working throughout with the background bosons. We have chain relations including the CS fields in this case and all unphysical fields(the bosons and the CS fields) disappear through successive replacement. Such an approach was pursued by Ezawa, Hotta and Iwazaki\cite{Ezawa92} in the framework of a nonrelativistic model. They obtained the hierarchy in which the filling factors are represented by continued fractions\cite{Haldane83}.

\section{A scalar field in the 1 space dimension}

The massless scalar field(Nambu-Goldstone boson) in the Schwinger\cite{Schwinger62} and the Thirring\cite{Thirring58} models was investigated by K.R. Ito\cite{KRIto75} and by Nakanishi\cite{Nakanishi75}. Ito gave an explicit model of the operators for this composite field. The annihilation operator is given by

\begin{equation}
  d^+(p^1)=\int \f{dq^1}{p^0}\{\theta(p^1) :\Psi_1^*(q^1)\Psi_1(p^1+q^1):
             + \theta(-p^1) :\Psi_2^*(q^1)\Psi_2(p^1+q^1):\},
\end{equation}
where
\[  \Psi(p^1)=u(p^1)a(p^1) +v(p^1)b^*(-p^1),  \]

\begin{equation}
  u(p^1)=\left(
	\begin{array}{c}\theta(p^1) \\ \theta(-p^1)
\end{array}
\right), \quad v(p^1)=\left(
	\begin{array}{c}\theta(-p^1) \\ \theta(p^1)
\end{array}
\right),                   \label{spn}
\end{equation}
 $\theta(x)$ is the step function and $a$($b$) is the annihilation operator for the original fermion(its antiparticle).\footnote{
We have modified the definition in the original paper. See the reference \cite{KRIto75} for further details.}
 The defined operator satisfies exactly the canonical commutation relations
\[ [ d^+(p^1), d^+(p'^1)^*]=\delta(p^1-p'^1), \quad \mbox{etc.}.  \]
This is due to the special charactor of the 1-dimensional 'spinor' (\ref{spn}).

\section{Discussion}

We have shown the canonical commutation relations for the 3-dimensional bound-state field. It is not persuadable to interpret the statistic property of the bound state by using the particle quantum mechanics. Instead, we should first conceive of the composite field consisting from the constituent fields as a nature being acquired by the space-time points. Translation of it into the language of the particle quantum mechanics may be that `the constituents loose their individuality in the bound state and behave as a quantum mechanical unity'\cite{Ito98}.

In the hierarchy of the compositeness, the Cooper pair of the super-conductor may be in the highest class. The annihilation operator for it is written as

\[ \Phi=\sum_{\blsmall{ k}}A(\bl{k})b_\uparrow(\bl{k})b_\downarrow(-\bl{k}). \]
We can show the canonical commutation relations by following the similar arguments leading to (\ref{cc1}) and (\ref{cc2}).

We have also shown the proper anyonic commutation relations for the composite state of two anyons. The phase exponents in the commutation relations of the constituent anyons add up to that of the composite anyon, as it is expected. This relation gives some criterion for anyons to be interpreted as (quasi)particles. We, further, investigated a class of the excitation(vortex) which can be identified with the anyonic quasiparticle but does not satisfy this criterion. We can say that the difference between the field theory and the many-particle quantum mechanics is more prominent in the 2 space dimension.

The theory of the composite boson in the FQHE was formulated in terms of the CS gauge field, which is a mathematical device representing the plane geometry of the system. We get the physical picture by eliminating this unphysical field. The (unphysical)background boson is converted into the physical anyon by virtue of the elimination. The commutation relations for the anyon field show the exclusion principle, except for the special case of the bosonic anyon. The generalized Laughlin ground state is formed as a consequence of the Fermi degeneracy.

 We saw that a composite field in the 1 space dimension satisfies the canonical commutation relations without restricting the Fock space of the original fermion.

%\section{References}

\section*{Acknowledgements}

The author would like to express his thanks to Professor Nakanishi for calling an attention to the 1-demensional case.
 He has been stimulated by discussions on BEC in the workshop "Thermo Field Dynamics" held at RIFP in the mid summers of 1999 and 2000. He would like to thank the organizers and participants with whom he enjoyed having discussions.

%\appendix

\end{document}